\documentclass[preprint,12pt]{elsarticle}

\usepackage{amssymb}
\usepackage{amsmath}

\usepackage{graphicx}
\usepackage{booktabs}
\usepackage{verbatim}
\usepackage{subcaption}
\usepackage{algorithm}
\usepackage{algpseudocode}

\usepackage[accsupp]{axessibility}  

\usepackage{enumitem}
\usepackage{multirow} 
\usepackage{setspace}

\usepackage{color, colortbl}

\newcommand{\rr}{\color{red}} 
\newcommand{\bb}{\color{black}}
\newcommand{\cc}{\color{cyan}} 
\newcommand{\ggr}{\color{green}}
\newcommand{\pp}{\color{magenta}}

\usepackage[pagebackref,breaklinks,colorlinks,citecolor=eccvblue]{hyperref}

\usepackage{orcidlink}

\journal{Information Fusion}

\begin{document}

\begin{frontmatter}



\title{Augmented Intelligence for Multimodal Virtual Biopsy in Breast Cancer Using Generative Artificial Intelligence}


\author[aff1]{Aurora Rofena}

\author[aff2]{Claudia Lucia Piccolo}

\author[aff2,aff3]{Bruno Beomonte Zobel}

\author[aff1,aff4]{Paolo Soda\corref{cor}}
\ead{p.soda@unicampus.it}

\author[aff1]{Valerio Guarrasi}


\affiliation[aff1]{organization={Unit of Computer Systems and Bioinformatics, Department of Engineering \\ Università Campus Bio-Medico di Roma},
            city={Rome},
            country={Italy}}

\affiliation[aff2]{organization={Department of Radiology, Fondazione Policlinico Campus Bio-Medico},
            city={Rome},
            country={Italy}}

\affiliation[aff3]{organization={Department of Radiology, Università Campus Bio-Medico di Roma},
            city={Rome},
            country={Italy}}

\affiliation[aff4]{organization={Department of Diagnostics and Intervention, Radiation Physics, Biomedical Engineering, Umeå University},
            city={Umeå},
            country={Sweden}}

\cortext[cor]{Correspondence: Paolo Soda}

\begin{abstract}
Full-Field Digital Mammography (FFDM) is the primary imaging modality for routine breast cancer screening; however, its effectiveness is limited in patients with dense breast tissue or fibrocystic conditions. Contrast-Enhanced Spectral Mammography (CESM), a second-level imaging technique, offers enhanced accuracy in tumor detection. Nonetheless, its application is restricted due to higher radiation exposure, the use of contrast agents, and limited accessibility. As a result, CESM is typically reserved for select cases, leaving many patients to rely solely on FFDM despite the superior diagnostic performance of CESM. While biopsy remains the gold standard for definitive diagnosis, it is an invasive procedure that can cause discomfort for patients.
We introduce a multimodal, multi-view deep learning approach for virtual biopsy, integrating FFDM and CESM modalities in craniocaudal and mediolateral oblique views to classify lesions as malignant or benign. To address the challenge of missing CESM data, we leverage generative artificial intelligence to impute CESM images from FFDM scans. 
Experimental results demonstrate that incorporating the CESM modality is crucial to enhance the performance of virtual biopsy. When real CESM data is missing, synthetic CESM images proved effective, outperforming the use of FFDM alone, particularly in multimodal configurations that combine FFDM and CESM modalities.
The proposed approach has the potential to improve diagnostic workflows, providing clinicians with augmented intelligence tools to improve diagnostic accuracy and patient care.
Additionally, as a contribution to the research community, we publicly release the dataset used in our experiments, facilitating further advancements in this field.
\end{abstract}




\end{frontmatter}





\section{Introduction}
\label{sec:intro}

Breast cancer has become a significant global health challenge, ranking as the most commonly diagnosed cancer worldwide by the end of 2020~\cite{bib:breast_cancer}.
That year, it affected 2.3 million women and caused nearly 685,000 deaths.
Projections estimate a more than 40\% increase in annual cases by 2040, reaching approximately 3 million, while related deaths are expected to rise by over 50\%, surpassing 1 million per year~\cite{bib:breast_cancer}.

Currently, Full-Field Digital Mammography (FFDM) is the primary imaging modality used in routine breast cancer screening.
FFDM, commonly referred to as standard mammography, uses low-energy X-rays to produce breast images typically from craniocaudal (CC) and mediolateral oblique (MLO) projections.
Despite FFDM being the standard mammography method, it has limitations, particularly in women with dense breast tissue, where overlapping structures can obscure lesions and contrast between normal and abnormal tissues.
Diagnostic accuracy can also be reduced in cases involving fibrocystic disease and post-treatment follow-ups after breast-conserving or adjuvant therapies \cite{bib:cesm-ffdm-mri}.

To overcome these limitations, Contrast Enhanced Spectral Mammography (CESM) has emerged as a valuable second-level imaging technique.
CESM uses an iodinated contrast agent to enhance the visualization of vascular structures associated with tumors, significantly improving the accuracy of tumor detection.
The technique involves dual-energy imaging during a single breast compression.
After administering the contrast agent, CC and MLO projections are captured, acquiring both low-energy (LE) and high-energy (HE) images.
The LE image is equivalent to a standard mammogram produced by FFDM, as validated by several studies \cite{bib:ffdm-le-1, bib:ffdm-le-2, bib:ffdm-le-3}, while the HE image, taken with a higher radiation dose, is not used directly for diagnostic purposes.
These LE and HE images are combined using a dual-energy weighted logarithmic subtraction technique, resulting in a recombined or dual-energy subtracted (DES) image that highlights areas of contrast uptake, improving lesion visibility.
For diagnostic evaluation, radiologists rely on both DES and LE images to ensure comprehensive assessment.

CESM is particularly valuable for patients with dense breast tissue or those at higher risk of cancer when FFDM alone is insufficient. \cite{bib:cesmbook}.
It offers diagnostic accuracy comparable to contrast-enhanced Magnetic Resonance Imaging (MRI) but with advantages such as reduced cost, shorter imaging times, and wider accessibility \cite{cesm-mri}.
Despite its benefits, CESM presents some drawbacks.
Patients are exposed to slightly higher radiation doses compared to FFDM, and there is a risk of allergic reactions to the contrast agent, which, although generally mild, can occasionally result in more severe complications such as contrast-induced nephropathy, shortness of breath, or facial swelling \cite{bib:cesmbook}.
Given these limitations and the role of CESM as a second-line imaging method, CESM is generally reserved for specific cases, leaving many patients reliant solely on FFDM despite the superior diagnostic capabilities of CESM.
Ultimately, confirming whether a lesion is malignant or benign often requires a biopsy, the gold standard for definitive diagnosis.
While highly reliable, biopsy is invasive, potentially uncomfortable, and time-consuming, with inherent risks and delays in treatment initiation.

In this context, Artificial Intelligence (AI) offers promising solutions to overcome existing limitations.
Specifically, we propose a multimodal, multi-view deep learning approach that integrates FFDM and CESM modalities in both CC and MLO views to enable a virtual biopsy for breast cancer, classifying lesions as malignant or benign.
Our approach addresses the issue of missing CESM data by employing generative AI techniques to synthesize CESM images from existing FFDM scans, ensuring continuity in diagnostic workflows. 
When both FFDM and CESM modalities are available, we process them through state-of-the-art classifiers to perform the virtual biopsy.
In cases where the CESM modality is unavailable, the synthesized CESM data enables the classification process to proceed, leveraging a multimodal framework enriched with generative insights.
Specifically, our contributions are:
\begin{itemize}
    \item Introduced a novel multimodal, multi-view deep learning approach for virtual biopsy, integrating FFDM and CESM modalities in CC and MLO views, and pioneering the use of generative AI to impute CESM images when unavailable.;
    \item Evaluated the impact of synthetic CESM images for virtual biopsy in both multimodal and unimodal contexts, comparing their performance to scenarios where only the FFDM modality is used without CESM data imputation.;
    \item Released an extended version of the CESM@UCBM dataset~\cite{bib:rofena2024deep}, provided in DICOM format and enriched with information from medical reports and biopsy results. 
\end{itemize}

\section{Related Works}
\label{sec:related_works}
In recent years, the application of AI, particularly deep learning, has revolutionized medical imaging by significantly improving diagnostic accuracy and enabling the development of automated systems for lesion detection and classification across various imaging modalities. 
Convolutional Neural Networks (CNNs) are increasingly prominent in medical image analysis \cite{bib:cnn_medical_image}.
CNN-based models, such as ResNet \cite{bib:resnet} and VGG \cite{bib:vgg}, have been widely adopted for the virtual biopsy classification task in a variety of medical imaging domains, such as Computed Tomography (CT) \cite{bib:ct_vb_resnet, bib:ct_vb_vgg, bib:ct_vb_vgg2}, MRI~\cite{bib:mri_virtual_biopsy_resnet50, bib:mri_virtual_biopsy_resnet50_2, bib:mri_virtual_biopsy_vgg16},
FFDM \cite{bib:resnet18_ffdm2, bib:resnet18_ben_vs_mal_ffdm, bib:mamm_vgg_resnet}, and CESM \cite{bib:cesm_vb_resnet} due to their ability to learn intricate spatial patterns in image data.

Multimodal learning, which integrates information from multiple imaging modalities, is gaining widespread adoption in medical imaging due to its potential to significantly improve diagnostic accuracy and performance~\cite{bib:multimodal_mif_review}.
For instance, combining MRI and PET enables the simultaneous analysis of both structural and functional information, proving highly effective in classifying Alzheimer's disease (AD) \cite{bib:multimodal_mri_pet}.
Similarly, integrating FDG-PET with CT allows for the concurrent evaluation of metabolic and anatomical data, offering significant advantages in lung cancer classification \cite{bib:multimodal_pet_ct}.
In breast cancer, multimodal learning models integrating mammogram and ultrasound images have shown improved accuracy in the virtual biopsy task compared to unimodal approaches \cite{bib:multimodal_mamm_us}.

A significant challenge in multimodal medical imaging is the issue of missing modalities.
In real-world clinical settings, the unavailability of certain imaging modalities is common, often due to cost constraints, patient-specific factors, or technical limitations \cite{bib:missing_modality}.
Despite its widespread occurrence, this issue has been examined in relatively few studies.
Managing incomplete multimodal data remains a significant challenge, with most existing methodologies opting to exclude subjects with incomplete data, thereby diminishing the sample size \cite{bib:review}.
Recently, multimodal image synthesis, referred to as data imputation, has gained attention as a promising solution, enabling the generation of missing modalities from the available data.
Current methods in medical imaging perform image-to-image translation by leveraging generative models~\cite{bib:missing_modality_2}, such as generative adversarial networks (GANs). 
Specifically, there are studies in the fields of MRI, PET, and CT that utilize GANs for data imputation, incorporating the generated modality within classification frameworks \cite{bib:mri_to_pet, bib:mri_to_pet_2, bib:ct_gan}.
However, to the best of our knowledge, no similar works exist in the context of FFDM and CESM.
Our research aims to bridge this gap by proposing a multimodal, multi-view deep learning approach that leverages FFDM and CESM modalities to perform a virtual biopsy of breast cancer.
The approach employs generative AI for data imputation, integrating synthetic CESM images generated from FFDM to address the challenge of missing modalities.

\section{Methods}
\label{sec:methods}

\begin{figure*}[h!] 
\includegraphics[width=\textwidth]{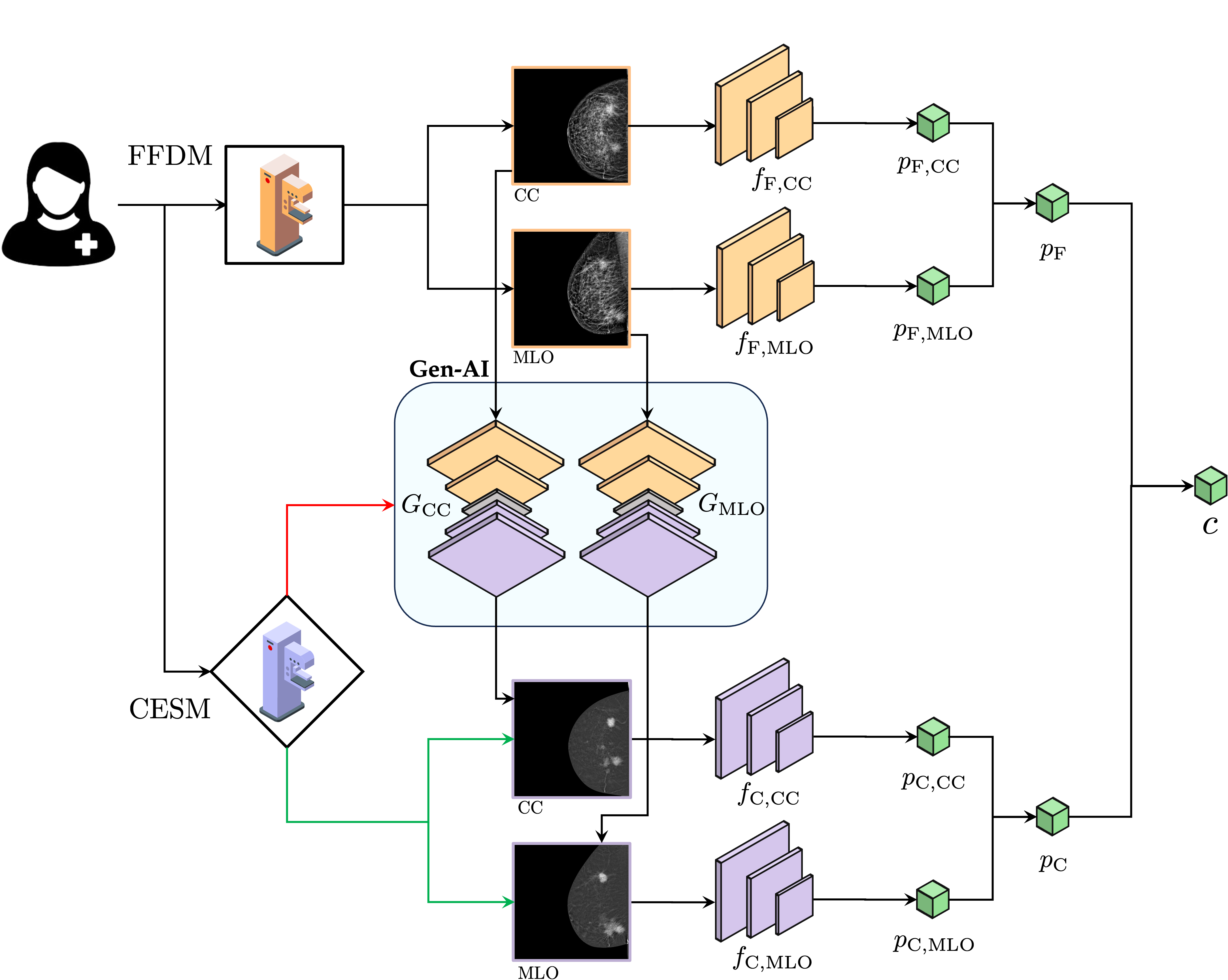}
    \caption{Schematic representation of the methodology.}
    \label{fig:methods}
\end{figure*}

We propose a multimodal, multi-view deep learning pipeline for performing breast cancer virtual biopsy, classifying lesions as malignant or benign.
This framework integrates FFDM and CESM as input modalities, utilizing both CC and MLO views for each modality.
If the CESM modality is unavailable, we employ a generative AI model to synthesize CESM images from the corresponding FFDM scans.
~\autoref{fig:methods} presents a schematic overview of the proposed methodology, with further details outlined in the current section.
This discussion focuses on the two key components of our approach: the virtual biopsy task, which involves classifying lesions as malignant or benign, and the generative process for synthesizing CESM images to address missing modalities.

\subsection{Virtual Biopsy}
\label{sec:virtual_biopsy}
The virtual biopsy task is formulated as a binary classification problem, where the objective is to determine whether a tumor lesion is malignant or benign.
Let $\mathcal{M} = \{ F, C \}$ denote the set of imaging modalities, where $F$ corresponds to FFDM and $C$ corresponds to CESM.
Let $\mathcal{V} = \{ \mathrm{CC}, \mathrm{MLO} \}$ denote the set of views.
For each modality $m \in \mathcal{M}$ and view $v \in \mathcal{V}$, we define the corresponding image as $X_{m,v}$.
Specifically:
\begin{itemize}
    \item $X_{F,\mathrm{CC}}$: is the FFDM image in CC view;
    \item $X_{F,\mathrm{MLO}}$: is the FFDM image in MLO view;
    \item $X_{C,\mathrm{CC}}$: is the CESM image in CC view;
    \item $X_{C,\mathrm{MLO}}$: is the CESM image in MLO view.
\end{itemize}
We employ the model $f_{m,v}$, tailored to a specific modality $m$ and view $v$, which takes as input the corresponding image $X_{m,v}$ and estimates the posterior probability of tumor lesion malignancy $p_{m,v}$.
This can be expressed as:
\begin{equation}
    p_{m,v} = f_{m,v}(X_{m,v}), \quad \text{with} \quad 0 \leq p_{m,v} \leq 1, \quad \forall m \in \mathcal{M}, \; \forall v \in \mathcal{V}.
\end{equation}
We aggregate the probabilities from the CC and MLO views within each modality using a late fusion function $\phi_{\mathcal{V}}$, which combines the outputs from both views to produce an unimodal, multi-view probability of malignancy $p_m$ for the given modality:
\begin{equation}
    p_m = \phi_{\mathcal{V}}\left( p_{m,\mathrm{CC}}, \, p_{m,\mathrm{MLO}}\right), \quad \text{with} \quad 0 \leq p_{m} \leq 1, \quad \forall m \in \mathcal{M}.
    \label{eq:fuse_views}
\end{equation}
Next, we combine the unimodal, multi-view probabilities using a multimodal late fusion function $\phi_{\mathcal{M}}$, which combines the outputs from both modalities, to obtain an overall multimodal, multi-view probability of malignancy $p$:
\begin{equation}
    p = \phi_{\mathcal{M}}\left( p_F, \, p_C \right), \quad \text{with} \quad 0 \leq p \leq 1.
    \label{eq:fuse_modalities}
\end{equation}
Finally, the predicted class $c$ is determined by a max membership rule.
Thus, if the overall probability of malignancy exceeds that of benignity, the model classifies the tumor lesion as malignant; otherwise, it classifies it as benign.

\subsection{Generative AI for Missing CESM Data}
\label{sec:gen_AI}

When the CESM modality is unavailable, we employ view-specific generative AI models, denoted as $G_v$ for each view $v \in \mathcal{V}$, to synthesize CESM images $\hat{X}_{C,v}$ from FFDM images $X_{F, v}$:
\begin{equation}
     G_v\left( X_{F,v} \right) = \hat{X}_{C,v}, \quad \forall v \in \mathcal{V}.
\end{equation}
Subsequently, for virtual biopsy, the synthetic CESM images $\hat{X}_{C,v}$ are processed by their corresponding $f_{\text{C}, v}$ in the same way as for the real CESM images $X_{\text{C},v}$, ensuring consistency in the analysis pipeline.

To summarize, Algorithm~\autoref{alg:pipeline} presents the pseudocode of the proposed multimodal, multi-view deep learning approach for virtual biopsy for a single patient.

\begin{algorithm}[ht]
\caption{Multimodal Multi-View Virtual Biopsy Pipeline}
\label{alg:pipeline}
\begin{algorithmic}[1]
\Require Modalities $\mathcal{M} = \{ F, C \}$, views $\mathcal{V} = \{ \mathrm{CC}, \mathrm{MLO} \}$, FFDM images $X_{F,v}$ for the patient
\Ensure Classification of tumor lesion for the patient

\For{each view $v \in \mathcal{V}$}
    \State $X_{F,v}$ \Comment{Load FFDM image}
    \If{CESM image $X_{C,v}$ is available}
        \State $X_{C,v}$ \Comment{Use real CESM image}
    \Else
        \State $\hat{X}_{C,v} \gets G_v\left( X_{F,v} \right)$ \Comment{Generate synthetic CESM image}
        \State $X_{C,v} \gets \hat{X}_{C,v}$
    \EndIf
\EndFor

\State \Comment{Virtual Biopsy}
\For{each modality $m \in \mathcal{M}$}
    \For{each view $v \in \mathcal{V}$}
        \State $p_{m,v} \gets f_{m,v}\left( X_{m,v} \right)$ \Comment{Estimate malignancy probability}
    \EndFor
    \State $p_m \gets \phi_{\mathcal{V}}\left( p_{m,\mathrm{CC}}, \, p_{m,\mathrm{MLO}} \right)$ \Comment{Fuse probabilities across views}
\EndFor
\State $p \gets \phi_{\mathcal{M}}\left( p_F, \, p_C \right)$ \Comment{Fuse probabilities across modalities}
\If{$p \geq 0.5$}
    \State \textbf{Output:} $c =$  \textbf{malignant}
\Else
    \State \textbf{Output:} $c =$ \textbf{benign}
\EndIf
\end{algorithmic}
\end{algorithm}

\section{Materials}
\label{sec:dataset}
\begin{figure*}[h!] 
    \includegraphics[width=\textwidth]{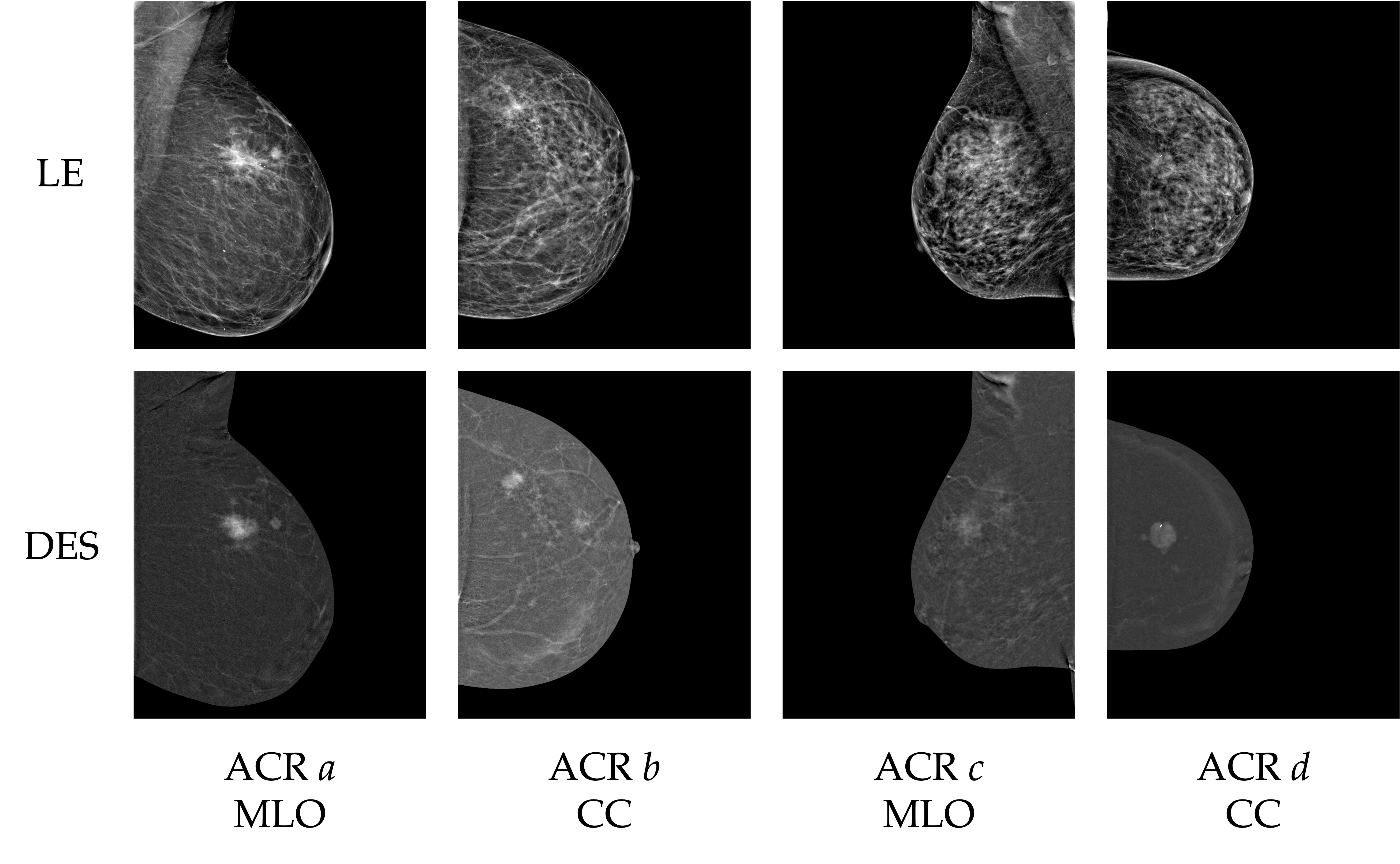}
    \caption{From left to right, pairs of LE (above) and DES images (below) with ACR categories a, b, c, d. 
    Breasts of ACR categories \textit{a} and \textit{c} are in the CC view, and breasts of ACR categories \textit{b} and \textit{d} are in the MLO view.}
    \label{fig:dataset}
\end{figure*}

In this study, we expanded the publicly available CESM@UCBM dataset \cite{bib:rofena2024deep} to create a comprehensive collection of CESM exams, and we make the extended version publicly available after anonymizing sensitive data.
The resulting dataset comprises CESM exams from 204 patients, ranging in age from 31 to 90 years, with an average age of 56.7 years and a standard deviation of 11.2 years.
All exams were conducted at the Fondazione Policlinico Universitario Campus Bio-Medico in Rome using the GE Healthcare Senographe Pristina system, with a total of 2278 images.
Among these, 1998 images have dimensions of $2850 \times 2396$ pixels, while the remaining 280 images are sized at $2294 \times 1916$ pixels.
The dataset comprises a balanced collection of LE and DES images.
In our experimental setup, we consider the DES images as representations of the CESM modality, due to the observable uptake of contrast medium in DES imaging.
Similarly, we utilize the LE images to represent the FFDM modality, based on their established equivalency to standard mammography.
~\autoref{tab:dataset} summarizes the main characteristics of the dataset.
It includes images captured in both CC and MLO views, with equal representation of FFDM and CESM modalities in the different views.
A subset of the dataset corresponds to late-phase acquisitions, which are captured after a delay to evaluate persistent contrast enhancement in breast tissue, in contrast with standard early-phase acquisitions.
Based on biopsy results, we identified images associated with malignant or benign findings.
Additionally, based on medical reports, we identified images associated with different breast density levels, categorized into one of the four types (\textit{a}, \textit{b}, \textit{c}, \textit{d}) as defined by the fifth edition of the Breast Imaging Reporting and Data System issued by the American College of Radiology (ACR) \cite{bib:acr}.
Specifically, ACR category \textit{a} indicates breasts that are almost entirely fatty, ACR category \textit{b} corresponds to breasts with scattered fibroglandular density, ACR category \textit{c}, represents heterogeneously dense breast tissue, and ACR category \textit{d} signifies extremely dense breasts.
\begin{table*}[h!] 
\begin{center}
\resizebox{0.7\textwidth}{!}{
\begin{tabular}{llc}
\toprule
\multirow{2}{*}{\textbf{General Info}} & Patients & 204 \\ & Total Images & 2278 \\ 
 \midrule
\multirow{2}{*}{\textbf{Image Modality}} & FFDM & 1139 images \\ 
 & CESM & 1139 images \\ 
\midrule
\multirow{2}{*}{\textbf{Image View}} & CC & 1146 images (573 FFDM + 573 CESM) \\ 
 & MLO & 1132 images (566 FFDM + 566 CESM) \\ 
\midrule
\multirow{2}{*}{\textbf{Phase Acquisition}} & Early & 1598 images \\ 
 & Late & 680 images \\ 
\midrule
\multirow{3}{*}{\textbf{Biopsy}} & Malignant & 622 images \\ 
 & Benign & 224 images \\ 
\midrule
\multirow{5}{*}{\textbf{ACR category}} & \textit{a} & 128 images \\ 
 & \textit{b} & 706 images \\ 
 & \textit{c} & 738 images \\ 
 & \textit{d} & 260 images \\ 
 & Not reported & 446 images \\ 
 \bottomrule
\end{tabular}}
\caption{Characteristics of the dataset.}
\label{tab:dataset}
\end{center}
\end{table*}
~\autoref{fig:dataset} compares 4 pairs of LE and DES images belonging to the dataset, which differ in the ACR category.
Breasts of ACR categories \textit{a} and \textit{c} are shown in the CC view, while breasts of ACR categories \textit{b} and \textit{d} are shown in the MLO view.

~\autoref{tab:number_images} presents the distribution of images utilized in our experimental study.
For the implementation of the virtual biopsy process outlined in~\autoref{sec:virtual_biopsy}, we excluded all late acquisitions from the dataset to maintain consistency in imaging conditions.
This reduced the number of FFDM and CESM images from 1139 each to 799.

Furthermore, the dataset was refined by focusing solely on images containing either malignant or benign lesions, thereby reducing the number to 230 FFDM and 230 CESM images.
This subset comprised 83 malignant and 32 benign cases for both CC and MLO views across the two modalities.
For the generative task, referred to as CESM Generation, we employed the entire dataset comprising 2278 images.
Specifically, this included 573 FFDM images and their corresponding 573 CESM images in the CC view, which were used to generate synthetic CESM images in the CC view from FFDM images in the same view.
Similarly, the dataset contained 566 FFDM images and 566 corresponding CESM images in the MLO view, which were utilized for generating CESM images in the MLO view from FFDM images in the corresponding view.

\begin{table}[]
\centering
\begin{tabular}{cccc}
\toprule
\textbf{Task} & \textbf{Images} & \textbf{Label} & \textbf{\# of Images} \\
\midrule
\multirow{8}{*}{Virtual Biopsy} 
    & \multirow{2}{*}{$X_{\text{F}, \text{CC}}$} & Malignant & 83 \\
    &                          & Benign & 32 \\ 
\cmidrule(lr){2-4}
    & \multirow{2}{*}{$X_{\text{F}, \text{MLO}}$} & Malignant & 83 \\
    &                           & Benign & 32 \\
\cmidrule(lr){2-4}
    & \multirow{2}{*}{$X_{\text{C}, \text{CC}}$} & Malignant & 83 \\
    &                          & Benign & 32 \\
\cmidrule(lr){2-4}
    & \multirow{2}{*}{$X_{\text{C}, \text{MLO}}$} & Malignant & 83 \\
    &                           & Benign & 32 \\
\midrule
\multirow{2}{*}{CESM Generation} 
    & $X_{\text{F}, \text{CC}}$ + $X_{\text{C}, \text{CC}}$ & - & 573 + 573 \\ 
\cmidrule(lr){2-4}
    & $X_{\text{F}, \text{MLO}}$ + $X_{\text{C}, \text{MLO}}$ & - & 566 + 566 \\
\bottomrule
\end{tabular}
\caption{Distribution of images in the tasks.}
\label{tab:number_images}
\end{table}

\subsection{Image Pre-processing}
\label{subsec:preProcessing}
We pre-processed all images to ensure data consistency and uniformity. 
Specifically, we squared the images by padding them with the average background value, followed by contrast stretching to enhance brightness and improve contrast.
Next, we normalized the pixel values to the range $[0, 1]$ and resized the images to $256 \times 256$, balancing computational efficiency with sufficient resolution.
Finally, we horizontally flipped all left breast images, ensuring that all images have a consistent orientation.

\section{Experimental Setup}
\label{sec:exp_setpu}
In this section, we outline the experimental setup of our study, detailing the training and evaluation processes for the classifiers and generative model used, as well as describing the experiments conducted to assess their robustness to missing modalities.

\subsection{Classifiers}
\label{sec:cnn}

In this work, to perform the virtual biopsy task, we evaluate three well-established CNN architectures: ResNet18~\cite{bib:resnet}, ResNet50~\cite{bib:resnet}, and VGG16~\cite{bib:vgg}.
These networks have demonstrated strong performance in classification tasks across various domains, including breast cancer virtual biopsy \cite{bib:virtual_biopsy_review}, making them suitable candidates for our multimodal, multi-view approach.

All experiments were conducted using a stratified 5-fold cross-validation to ensure that each fold contained a representative distribution of the target labels.
We split the dataset into training, validation, and test sets in proportions of 80\%, 10\%, and 10\%, respectively.
To avoid data leakage, we allocated all images from the same patient to the same set.
To prevent overfitting, we applied random data augmentation to the training set, including vertical and horizontal shift (up to $\pm$10\% of the original dimension), zoom (up to $\pm$10\%), and rotation (up to $\pm$15°).

\subsubsection{Training}
We trained each CNN for up to 300 epochs, employing an early stopping criterion based on validation loss, with a patience of 50 epochs, following an initial 50-epoch warm-up phase.
We used the Cross-Entropy loss function and the Adam with an initial learning rate of $10^{-3}$, $\beta=0.9$, and momentum of 1.
If the validation loss did not improve for 10 consecutive epochs, we reduced the learning rate by a factor of 0.1.
A weight decay of $10^{-5}$ was applied to enhance generalization and prevent overfitting.
We did not further investigate any other hyperparameter configuration since their tuning is out of the scope of this manuscript.
Nevertheless, the ``No Free Lunch” Theorem for optimization asserts that no single set of hyperparameters can universally optimize model performance across all datasets \cite{bib:no_free_lunch}.

\subsubsection{Evaluation}
\label{sec:eval}

In the inference phase, we leverage the trained classifiers $f_{m,v}$ to conduct the virtual biopsy on the corresponding test set images $X_{m,v}$, classifying lesions as either malignant or benign.
Each classifier $f_{m, v}$ outputs a malignancy probability $p_{m, v}$, specific for the modality $m$ and view $v$.
Then, based on a max membership rule, it predicts the lesion class $c_{m, v}$ as malignant or benign.
To evaluate the performance of the classifiers, we compute the following metrics based on the predicted classes and the ground truth:
\begin{itemize}
    \item Area Under the ROC Curve (AUC): This metric reflects the model's ability to distinguish between classes, summarizing the trade-off between true positive and false positive rates across different classification thresholds. A higher AUC indicates better discriminative performance.
    \item Geometric Mean (G-mean): The G-mean captures the balance between Recall and Specificity, being particularly useful for imbalanced datasets.
    It is computed as:
    \begin{equation}
    \text{G-mean} = \sqrt{\text{Recall} \times \text{Specificity}}
    \end{equation}
    where Recall and Specificity measure the proportion of actual positive and negative cases correctly identified, respectively.
    A high G-mean ensures balanced performance across both classes.
    \item Matthews Correlation Coefficient (MCC): The MCC considers True Positives (TP), False Positives (FP), True Negatives (TN), and False Negatives (FN), providing a balanced evaluation for imbalanced datasets. It returns a value between -1 and 1, with 1 indicating perfect prediction, 0 random prediction, and -1 total disagreement between predicted and actual classes. MCC is calculated as:
    \begin{equation}
    \text{MCC} = \frac{(\text{TP} \times \text{TN}) - (\text{FP} \times \text{FN})}{\sqrt{(\text{TP} + \text{FP})(\text{TP} + \text{FN})(\text{TN} + \text{FP})(\text{TN} + \text{FN})}}
    \end{equation}
\end{itemize}

\subsubsection{Fusion Strategy}
\label{sec:late_fusion_strategy}
As discussed in~\autoref{sec:virtual_biopsy}, we use a late fusion strategy to achieve a multimodal, multi-view classification for each lesion.
This process involves two stages of integration.
Initially, the fusion function $\phi_{\mathcal{V}}$ aggregates the probabilities of malignancy across views within each modality, obtaining the unimodal, multi-view probabilities of malignancy.
Then, the fusion function $\phi_{\mathcal{M}}$ integrates the unimodal, multi-view probabilities of malignancy, obtaining an overall multimodal, multi-view probability of malignancy.
For both fusion functions, we utilize a weighted average strategy, where probabilities are averaged using weights given by the MCC values computed from the validation set.
This approach is motivated by the ability of MCC to address class imbalance effectively, providing a robust measure of model performance, even in datasets with disproportionate class distributions \cite{mcc}.
Thus, for each modality $m \in \mathcal{M}$, we compute the unimodal, multi-view probability of malignancy $p_{m}$ as follows:
\begin{equation}
p_m = \frac{\sum_{v \in \{\text{CC}, \text{MLO}\}} \left(p_{m,v} \cdot \text{MCC}_{m,v}\right)}{\sum_{v \in \{\text{CC}, \text{MLO}\}} \text{MCC}_{m,v}}
\end{equation}
where $p_{m,v}$ represents the probability of malignancy for lesions in images from modality $m$ and view $v$, while $\text{MCC}_{m, v}$ represents the MCC value computed on the validation set for images of modality $m$ and view $v$, based on the predicted class $c_{m, v}$ and the ground truth.
For each modality, we achieve an unimodal, multi-view classification by predicting the class $c_m$ based on the $p_m$ value.
Subsequently, we aggregate the unimodal, multi-view probabilities, using the same weighted strategy to compute the multimodal, multi-view probability of malignancy $p$ for each lesion, as follows:
\begin{equation}
p = \frac{\sum_{m \in \{F, C\}} p_m \cdot \text{MCC}_m}{\sum_{m \in \{F, C\}} \text{MCC}_m}
\end{equation}
where $\text{MCC}_m$ represents the MCC value computed on the validation set for images of modality $m$, based on the predicted class $c_m$ and the ground truth.
We achieve a multimodal, multi-view classification by predicting the class $c$ based on the $p$ value.

\subsection{Generative model}
\label{sec:gen_model}
We utilize the CycleGAN \cite{bib:cyclegan} as the generative model due to its demonstrated effectiveness in generating CESM images from FFDM images \cite{bib:rofena2024deep}.

The experiments were conducted using a 5-fold cross-validation strategy.
The dataset was divided into training, validation, and test sets in proportions of 80\%, 10\%, and 10\%, respectively, ensuring consistency with the splits used for the classification task in the virtual biopsy.
To prevent overfitting, we applied random data augmentation to the training set, including vertical and horizontal shift (up to $\pm$10\% of the original dimension), zoom (up to $\pm$10\%), and rotation (up to $\pm$15°).

\subsubsection{Training}
Given the set of imaging views $\mathcal{V} = \{ CC, MLO \}$, we trained the CycleGAN to generate CESM images $\hat{X}_{\text{C}, v}$ from FFDM images $X_{\text{F}, v}$, where $v \in \mathcal{V}$.
The training process involves two parallel pathways: one for generating CESM images in the CC view ($\hat{X}_{\text{C},\text{CC}}$) from FFDM images in the same view ($X_{\text{F},\textbf{CC}}$), and another for generating CESM images in the MLO view ($\hat{X}_{\text{C},\text{MLO}}$) from FFDM images in the same view ($X_{\text{F},\text{MLO}}$).
Our objective is to train the CycleGAN to synthesize CESM images, $\hat{X}_{\text{C},\text{CC}}$ and $\hat{X}_{\text{C},\text{MLO}}$, that closely resemble the corresponding real CESM images, $X_{\text{C},\text{CC}}$ and $X_{\text{C},\text{MLO}}$. 
This can be formalized as follows:
\begin{equation}
G_v(X_{\text{F}, v})~=~\hat{X}_{\text{C}, v}~\approx~X_{\text{C}, v} \quad \forall v \in \mathcal{V}
\end{equation}

After a pre-training on the public dataset described in \cite{bib:PublicDataset}, we fine-tuned the CycleGAN separately for the CC and MLO views using the dataset introduced in~\autoref{sec:dataset}, which had been preprocessed as outlined in~\autoref{subsec:preProcessing}.
Both training processes ran for up to 200 epochs, employing an early stopping criterion based on validation loss, with patience of 50 epochs, following an initial 50-epoch warm-up phase.
Although CycleGAN can typically operate with unpaired datasets, our study used paired data to compute losses by directly comparing the synthesized CESM images $\hat{X}_{\text{C}, v}$ with the corresponding real CESM images $X_{\text{C}, v}$.
To ensure effective and coherent image-to-image translation, we employ three complementary loss functions:
\begin{itemize}
    \item Adversarial Loss: encourages the generations of images indistinguishable from real target domain images, enhancing the realism of the synthesized outputs;
    \item Cycle Consistency Loss: ensures that translating an image to the target domain and then back to the original domain preserves its structure;
    \item Identity Loss: helps maintain the original image characteristics when the input already belongs to the target domain, preventing unnecessary changes.
\end{itemize}
We used Mean Squared Error for the adversarial loss, and $L1$ loss for both cycle consistency and identity mapping losses, with weighting factor of $\lambda_1 = 10$ and $\lambda_2 = 5$ respectively.
We optimized the generator and discriminator networks using the Adam optimizer, with a learning rate of $10^{-5}$, weight decay of $10^{-5}$, a beta of $0.5$, and momentum set to $1$.
We did not further investigate any other hyperparameter configuration since their tuning is out of the scope of this manuscript.
Nevertheless, the ``No Free Lunch” Theorem for optimization asserts that no single set of hyperparameters can universally optimize model performance across all datasets.

\subsubsection{Evaluation}
The quality of image generation is quantitatively assessed on the test set, calculating three metrics between the synthetic CESM images $\hat{X}_{\text{C}, v}$ and the target CESM images $X_{\text{C}, v}$.
For simplicity, we will refer to synthetic CESM images as $\hat{y}$ and real CESM images as $y$ throughout this section.
The selected metrics are Mean Squared Error (MSE), Peak-Signal-to-Noise Ratio (PSNR), and Structural Similarity Index (SSIM).

The MSE quantifies the mean squared difference between the pixel values of the target image $y$ and the reconstructed image $\hat{y}$, thus is formulated as:
\begin{equation}
    \text{MSE}(y,\hat{y}) = \frac{1}{kh}\sum_{i=1}^{k}\sum_{j=1}^{h}(y_{ij}-\hat{y}_{ij})^2
   \label{eq: mse}
\end{equation}
where $k$ and $h$ are the number of rows and columns in the images, respectively, and $y_{ij}$ and $\hat{y}_{ij}$ represent the pixels elements at the
$i$-th row and $j$-th column of $y$ and $\hat{y}$, respectively.
It varies in the range $[0, \infty]$, with lower values indicating higher quality of the reconstructed image.

The PSNR is defined as the ratio of the maximum possible power of a signal to the power of the noise that affects the signal.
In this context, the signal represents the target image, while the noise corresponds to the error introduced during its reconstruction.
The PSNR is expressed in decibels (dB), with a PSNR value of 30 dB considered as excellent quality, 27 dB as good quality, 24 dB as poor quality, and 21 dB as bad quality~\cite{psnr_quality}.
It is commonly expressed as a function of the MSE as follows:
\begin{equation}\begin{split}
    \text{PSNR}(y, \hat{y}) = 10 \cdot \log_{10} \left( \frac{\text{max}^2(y)}{\text{MSE}(y, \hat{y})} \right)
   \label{eq: psnr}
\end{split}\end{equation}

The SSIM~\cite{ssim} measures the similarity between two images by comparing their luminance, contrast, and structure.
It is defined as follows:
\begin{equation}
    \text{SSIM}(y, \hat{y}) = \frac{{(2\mu_y\mu_{\hat{y}} + B_1)(2\sigma_{y\hat{y}} + B_2)}}{{(\mu_y^2 + \mu_{\hat{y}}^2 + B_1)(\sigma_y^2 + \sigma_{\hat{y}}^2 + B_2)}}
    \label{eq:ssim}
\end{equation}
where $\mu_y$ and $\mu_{\hat{y}}$ represent the means of the images $y$ and $\hat{y}$ respectively, while $\sigma_y$ and $\sigma_{\hat{y}}$ represent their standard errors. $B_1$ and $B_2$ are small constants used for stabilization.
The SSIM varies in the range $[0, 1]$, with higher values signifying greater similarity between the synthetic and target images~\cite{ssim}.

\subsection{Missing Modality Robustness}

In clinical practice, it is common for certain patients to undergo only FFDM without the additional CESM examination.
To mimic this real-world limitation and evaluate the robustness of the virtual biopsy under missing data conditions, we simulate scenarios with varying proportions of missing CESM images among patients.
In these scenarios, we repeat the inference process using imputed CESM data $\hat{X}_{\text{C}, v}$ generated with CycleGAN to replace the missing CESM inputs $X_{\text{C},v}$ for the respective classifier $f_{\text{C}, v}$.
We assess the effect of different proportions of real and synthetic CESM images on virtual biopsy performance, by testing datasets containing $n\%$ synthetic and $(100-n)\%$ real CESM images in the CC and MLO views, with $n$ ranging from 10 to 100 in increments of 10.
For each $n$ value, we randomly select patients requiring synthetic CESM images, repeating the sampling process 10 times to ensure statistical robustness.
For each sampling
at a given $n$ value, we perform the virtual biopsy process.
We evaluate the classification performance using AUC, G-mean, and MCC and obtain a comprehensive performance for each value of $n$ by averaging these metrics across the $10$ sampling.

To evaluate the impact of AI-generated CESM images $\hat{X}_{\text{C},v}$, we compare the virtual biopsy performance obtained with real FFDM images $X_{\text{F}, v}$ and varying proportions of synthetic CESM images $\hat{X}_{\text{C},v}$ against two benchmarks: (1) performance with both real FFDM images $X_{\text{F}, v}$ and real CESM images $X_{\text{C}, v}$, and (2) performance with only real FFDM images $X_{\text{F}, v}$.
The performance with real FFDM and real CESM images establishes the upper bound, representing the ideal scenario where all imaging modalities are available.
In contrast, the performance with only real FFDM images defines the lower bound, reflecting the worst-case scenario where CESM is not performed, and virtual biopsy relies solely on FFDM.
Since our approach generates CESM images when unavailable and integrates them into the virtual biopsy workflow, the FFDM-only scenario is particularly relevant as the baseline for evaluating the effectiveness of our method.

\section{Results and Discussion}

In this section, we first present and discuss the results achieved from the generation of synthetic CESM images, and then we move to the results of the virtual biopsy task.

\subsection{Generative AI for Missing CESM Data}
\begin{figure*}[h!] 
    \includegraphics[width=\textwidth]{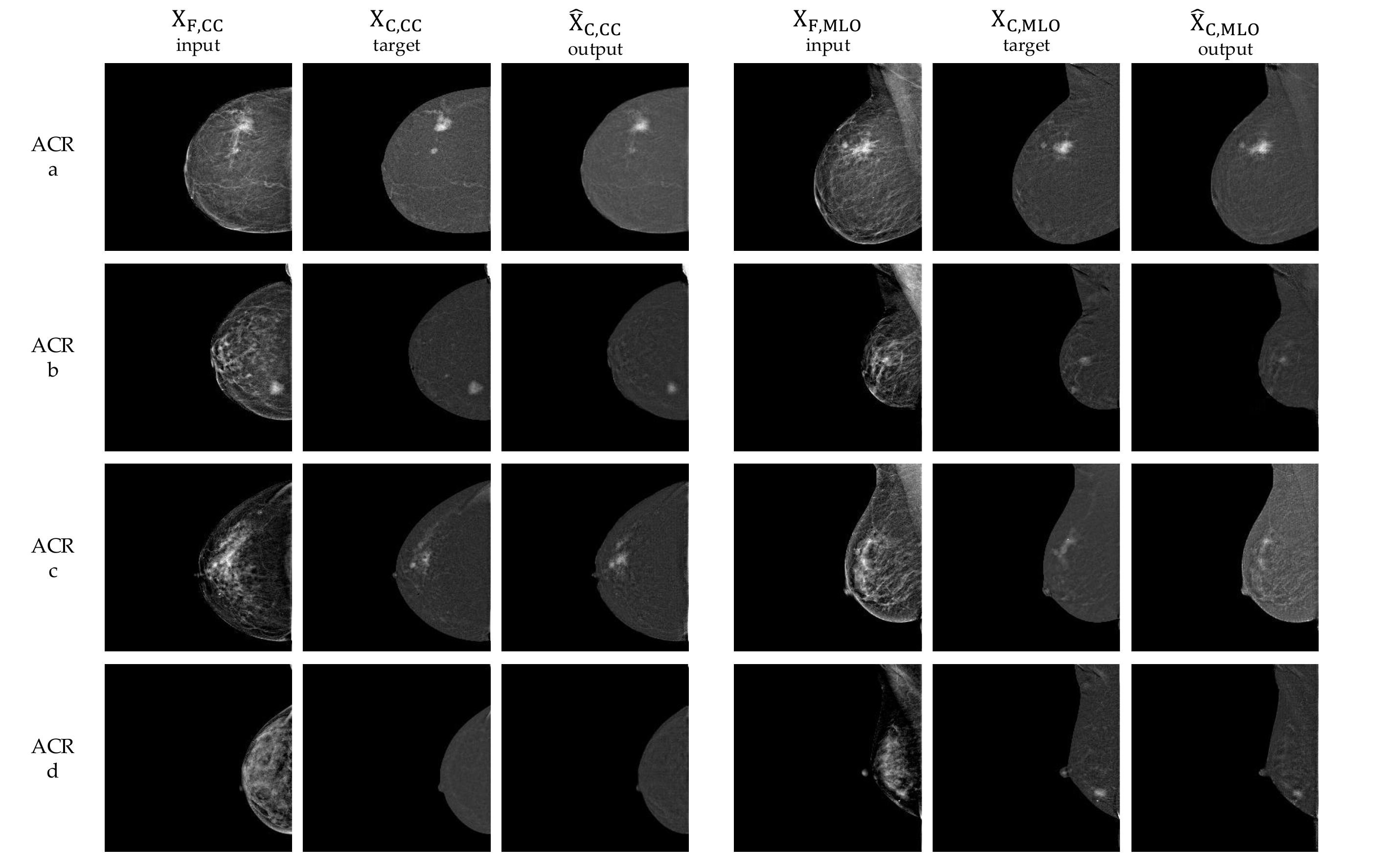}
    \caption{Representative cases of synthesized CESM images generated for both CC and MLO views, across all the ACR breast density categories. For each view and each ACR breast density category, FFDM input images, CESM target images, and CESM output images are shown.}
    \label{fig:examples}
\end{figure*}

Here we present the evaluation of CycleGAN's capability to generate CESM images from FFDM inputs for both CC and MLO views.
~\autoref{tab:cyclegan} summarizes the quantitative analysis, including the MSE, PSNR, and SSIM metrics computed on the test sets across all folds.
The lower MSE, higher PSNR, and higher SSIM observed for the CC view indicate that the model achieves better performance for this view, likely due to inherent challenges associated with the MLO view \cite{bib:cc_vs_mlo}.
Despite this, the MLO results remain within acceptable ranges, demonstrating the model's robustness.
Notably, PSNR values exceeding 24 dB and SSIM scores above 0.8 across both views validate CycleGAN's ability to generate high-quality CESM images that closely resemble the target images.
~\autoref{fig:examples} illustrates representative examples of synthesized CESM images for both views, across varying breast densities, classified as ACR \textit{a}, \textit{b}, \textit{c}, and \textit{d}.
For each density category, the FFDM input ($X_{\text{F}, \text{CC}}$ or $X_{\text{F}, \text{MLO}}$), the CESM target ($X_{\text{C}, \text{CC}}$ or $X_{\text{C}, \text{MLO}}$), and the synthesized CESM output ($\hat{X}_{\text{C},\text{CC}}$ or $\hat{X}_{\text{C},\text{MLO}}$) are displayed.
These examples visually demonstrate the model's ability to reconstruct CESM features with high fidelity, effectively capturing critical anatomical and pathological details across varying breast densities.
Together, the quantitative metrics and visual examples validate the effectiveness of CycleGAN in generating high-quality CESM images from FFDM inputs for both CC and MLO views.
This capability enables its application in the virtual biopsy task when CESM data is unavailable.

\begin{table}[]
\centering
\renewcommand{\arraystretch}{1.3} 
\setlength{\tabcolsep}{10pt} 
\begin{tabular}{ccccc}
\hline
\textbf{View} & \textbf{MSE} & \textbf{PSNR} & \textbf{SSIM} \\ \hline
CC & 0.0030 \text{\tiny ± 0.0003} & 27.7069 \text{\tiny ± 0.5238} & 0.8669 \text{\tiny ± 0.0143} \\ 
MLO & 0.0039 \text{\tiny ± 0.0008} & 26.5994 \text{\tiny ± 0.9930} & 0.8303 \text{\tiny ± 0.0175} \\ \hline
\end{tabular}
\caption{Quantitative analysis of CESM image generation from FFDM. Each tabular shows the mean value together with the standard error calculated across the 5 cross-validation folds.}
\label{tab:cyclegan}
\end{table}

\subsection{Virtual Biopsy}
\begin{figure}[h!]
    \centering
    {\includegraphics[width=1\textwidth]{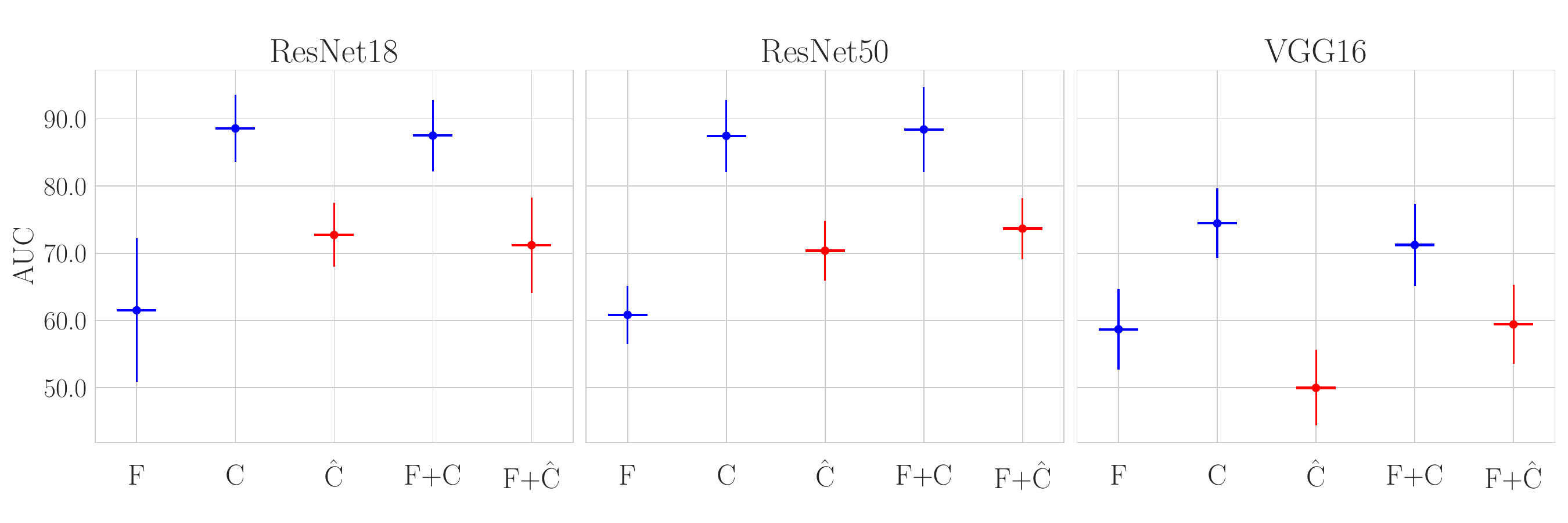}}
    
    \vspace{0.5cm}
    
    {\includegraphics[width=1\textwidth]{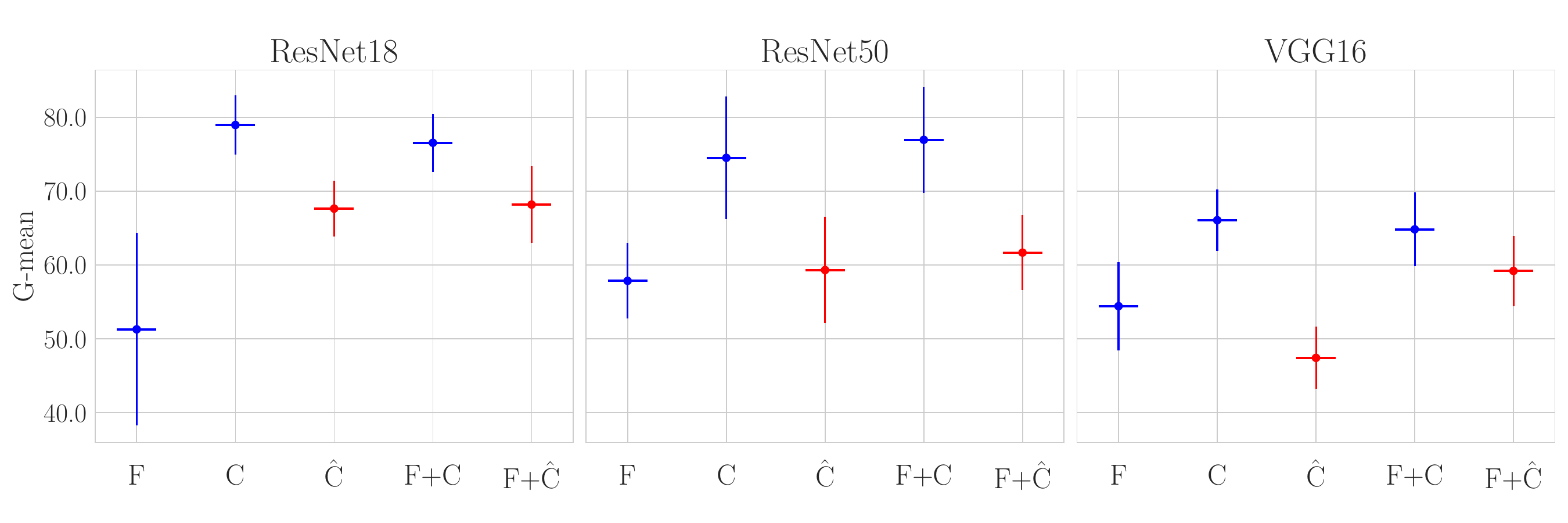}}
    
    \vspace{0.5cm}
    
    {\includegraphics[width=1\textwidth]{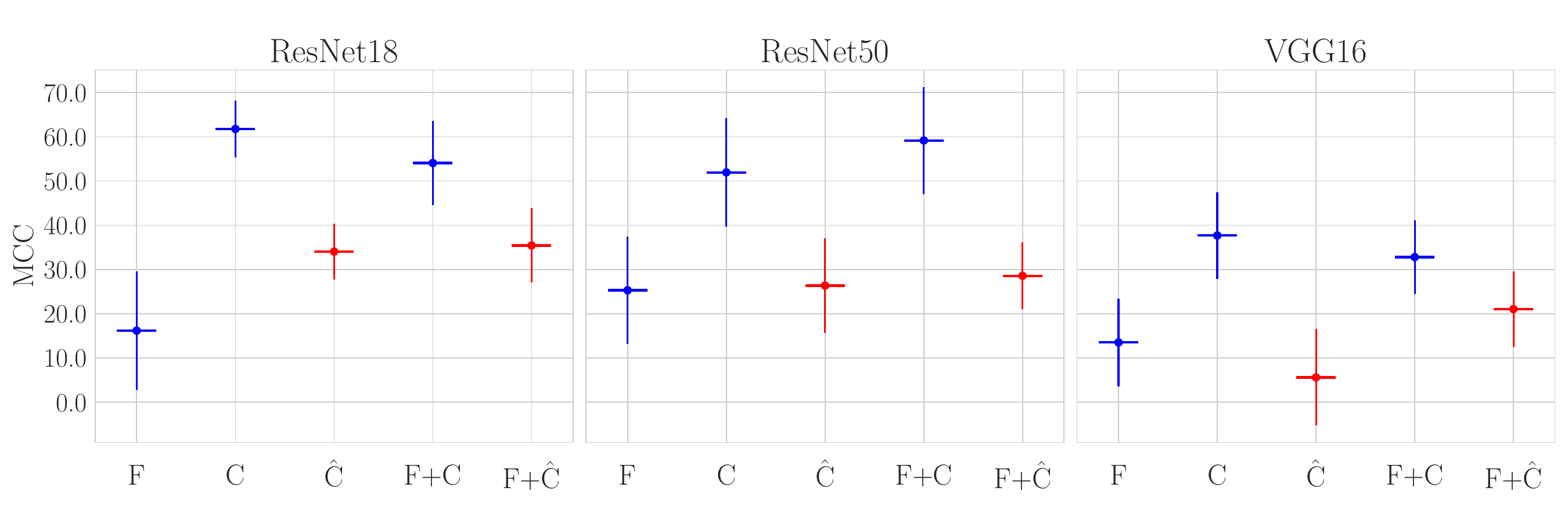}}
    
    \caption{AUC, G-mean, and MCC metrics across five experimental setups: classification using only FFDM modality (F), real CESM modality (C), synthetic CESM modality ($\hat{\text{C}}$), combined FFDM and CESM (F + C), and combined FFDM and synthetic CESM (F + $\hat{\text{C}}$). Results are reported for ResNet18, ResNet50, and VGG16 classifiers.}
    \label{fig:results}
\end{figure}

\begin{figure*}[h!] 
    \includegraphics[width=\textwidth]{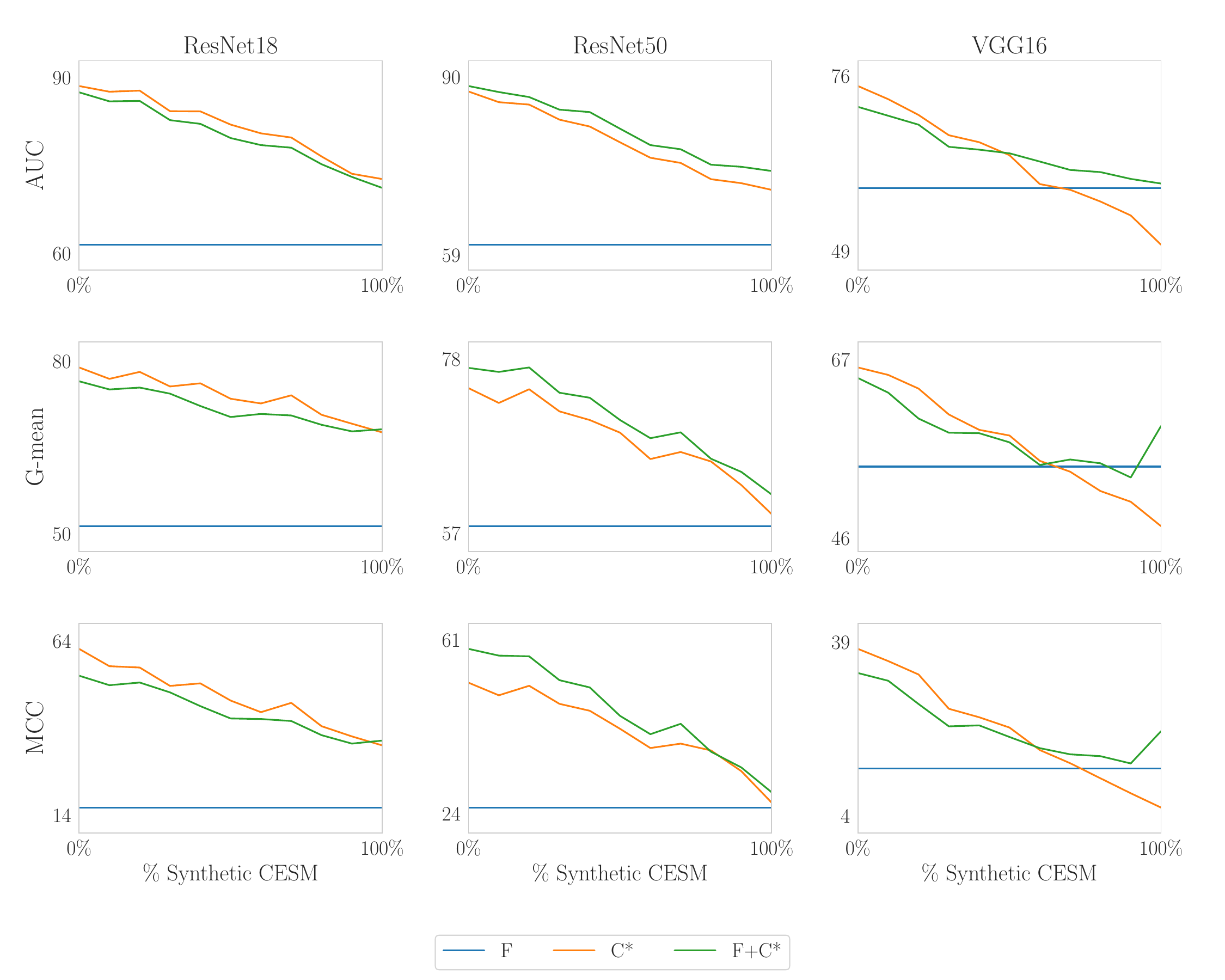}
    \caption{Robustness analysis of classification performance (AUC, G-mean, and MCC) across three experimental setups: classification using only FFDM modality (F), CESM modality with the proportion of synthetic CESM images varying from 0\% to 100\%(C*), combined FFDM and CESM with the proportion of synthetic CESM images varying from 0\% to 100\%(F+C*).}
    \label{fig:robustness}
\end{figure*}

Let us now turn the attention to the results of our multimodal, multi-view classification pipeline designed to perform the virtual biopsy.
~\autoref{fig:results} and ~\autoref{fig:robustness} provide a graphical overview of the experimental outcomes. 
The former figure  displays for each classifier the mean values and standard errors of the classification metrics across the following experimental settings:
\begin{itemize}
    \item F: virtual biopsy performed using only FFDM modality, combining CC and MLO views;
    \item C: virtual biopsy performed using only CESM modality, combining CC and MLO views;
    \item $\hat{\text{C}}$: virtual biopsy performed using only synthetic CESM modality generated by CycleGAN, combining CC and MLO views;
    \item F+C: virtual biopsy combining the results of F and C;
    \item F+$\hat{\text{C}}$: virtual biopsy combining the results of F and $\hat{\text{C}}$.
\end{itemize}

~\autoref{fig:robustness} investigates the robustness of the classification framework by varying the percentage of synthetic CESM images used.
It evaluates the performance of the F, C*, and F+C* experimental settings across the different classifiers. 
Specifically:
\begin{itemize}
    \item As for~\autoref{fig:results}, the F setting represents virtual biopsy using only FFDM modality, combining CC and MLO views, and is inherently unaffected by the proportion of synthetic CESM images;
    \item The C* setting corresponds to experiments where the CESM modality is used, with the proportion of synthetic CESM images varying from 0\% to 100\% in increments of 10\%. A 0\% proportion corresponds to the C setting (real CESM only), while a 100\% proportion corresponds to the $\hat{\text{C}}$ setting (synthetic CESM only).
    \item The F+C* setting involves combining the FFDM and CESM modalities, where the CESM modality includes a varying percentage of synthetic data. At 0\% synthetic CESM, this setting corresponds to F+C, and at 100\% synthetic CESM, it corresponds to F+$\hat{\text{C}}$.
\end{itemize}

From both figures, it is evident that for both ResNet models and VGG16, the experimental settings C and F+C consistently achieve higher values of AUC, G-mean, and MCC compared to the F setting.
This finding highlights the critical role of the CESM modality in enhancing the virtual biopsy process, aligning with the study's objective of exploring the use of synthetic CESM images as a viable alternative when real CESM images are unavailable.
However, as shown in~\autoref{fig:robustness}, the performance of both the C* and F+C* settings declines as the proportion of synthetic CESM data increases.
Despite this decrease, the results demonstrate that even synthetic CESM images positively contribute to virtual biopsy when compared to the F setting, which uses only the FFDM modality.
Specifically, for ResNet18, both C* and F+C* consistently outperform the F setting, including the $\hat{\text{C}}$ and F+$\hat{\text{C}}$ scenarios, where 100\% of synthetic CESM images are involved.
These results support the utility of synthetic CESM data in improving virtual biopsy when the real CESM modality is unavailable.
Similarly, for ResNet50, the C* and F+C* settings outperform F, with the F+C* setting providing a slight improvement over the C* setting, further highlighting the benefits of the multimodal approach.
On the other hand, for VGG16, the C* setting demonstrates a performance decline relative to F when the proportion of synthetic CESM data exceeds 70\%.
However, the F+C* setting mitigates this decline, delivering improved performance compared to both the C* and F settings, even in the F+$\hat{\text{C}}$ scenario.
This reinforces the advantage of the multimodal approach, confirming that integrating synthetic CESM images is preferable to relying solely on FFDM.
The numerical values of metrics for the F, C, $\hat{\text{C}}$, F+C, and F+$\hat{\text{C}}$ experimental settings, are provided in~\autoref{tab:results} in the appendix. 
These values are expressed on a $[0, 100]$ scale for AUC and G-mean, and on a $[-100, 100]$ scale for MCC, across all classifiers.
As a further analysis,~\autoref{tab:acr} shows the AUC, G-mean, and MCC metrics calculated for test set images grouped according to breast density, as defined by the ACR categories \textit{b}, \textit{c}, or \textit{d}.
We compare the performance of the virtual biopsy across three experimental settings: F, F+C, and F+$\hat{\text{C}}$.
For clarity, results for the C and $\hat{\text{C}}$ settings are excluded, as previous analyses have demonstrated the advantages of the multimodal approach. 
This investigation aims to assess whether the F+$\hat{\text{C}}$ configuration, which integrates synthetic CESM images, consistently outperforms the F setting, reliant solely on FFDM images, particularly for dense breasts under the \textit{c} and \textit{d} categories, where tumor masses are often obscured by surrounding tissue.
Together with the performance of the three models, we report their average performance in the ``AVG” column. 
We use this average performance as the basis for evaluation since our focus is not on identifying the top-performing individual classifier for virtual biopsy but rather on demonstrating the efficacy of the multimodal approach, also when synthetic CESM images are utilized.
As expected, the F+C setting consistently achieves the best performance across all breast density categories.
Thus, the subsequent discussion focuses on the comparison between the F and F+$\hat{\text{C}}$ settings.
Notably, for category \textit{a}, where all test images belong to the same class, AUC cannot be computed, and while MCC could technically be calculated, it would always yield a value of 0, rendering it uninformative.
Consequently, G-mean is the only meaningful metric in this scenario, and the results clearly show that the F+$\hat{\text{C}}$ configuration surpasses the F setting. 
Similarly, for the more challenging categories, represented by ACR \textit{c} and \textit{d}, all metrics (AUC, G-mean, and MCC) exhibit superior performance under the F+$\hat{\text{C}}$ setting, confirming the added value of synthetic CESM images for the virtual biopsy.
However, this trend is not observed for ACR category \textit{b}, where the F+$\hat{\text{C}}$ setting shows a slight decline across all metrics compared to the F setting. 
Despite this exception, the consistent gains observed for dense breast categories highlight the effectiveness of incorporating synthetic CESM in enhancing virtual biopsy when real CESM are unavailable.

\begin{table}[htbp]
\centering
\renewcommand{\arraystretch}{1.3} 
\setlength{\tabcolsep}{10pt} 
\label{tab:combined_metrics}
\resizebox{\textwidth}{!}{ 
\begin{tabular}{cclccc|c}
\hline
\textbf{Metric} & \textbf{ACR} & \textbf{Setting} & \textbf{ResNet18} & \textbf{ResNet50} & \textbf{VGG16} & \textbf{AVG} \\ \hline
\multirow{12}{*}{\textbf{AUC}} 
 &  & F           & 68.33 \text{\tiny ± 21.14}   & 80.83 \text{\tiny ± 6.48}    & 88.34 \text{\tiny ± 7.04}    & 79.17 \text{\tiny ± 10.11}   \\
 &  b & F + C       & 95.00 \text{\tiny ± 4.47}    & 96.94 \text{\tiny ± 1.59}    & 88.33 \text{\tiny ± 6.14}    & 93.42 \text{\tiny ± 4.52}    \\
 &   & F + $\hat{\text{C}}$ & 74.72 \text{\tiny ± 7.99}    & 76.39 \text{\tiny ± 7.14}    & 83.89 \text{\tiny ± 6.31}    & 78.33 \text{\tiny ± 4.88}    \\ \cline{2-7}
 &  & F           & 65.64 \text{\tiny ± 11.56}   & 53.17 \text{\tiny ± 11.43}   & 43.56 \text{\tiny ± 15.25}   & 54.12 \text{\tiny ± 11.07}   \\
 &  c & F + C       & 85.70 \text{\tiny ± 6.87}    & 78.14 \text{\tiny ± 13.64}   & 61.57 \text{\tiny ± 8.43}    & 75.14 \text{\tiny ± 12.34}   \\
 &   & F + $\hat{\text{C}}$ & 61.16 \text{\tiny ± 16.73}   & 65.77 \text{\tiny ± 11.94}   & 54.52 \text{\tiny ± 18.03}   & 60.48 \text{\tiny ± 5.66}    \\ \cline{2-7}
 &  & F           & 37.50 \text{\tiny ± 7.91}    & 81.25 \text{\tiny ± 11.86}   & 68.75 \text{\tiny ± 19.76}   & 62.50 \text{\tiny ± 22.53}   \\
 &  d & F + C       & 81.25 \text{\tiny ± 11.86}   & 100.00 \text{\tiny ± 0.00}   & 87.50 \text{\tiny ± 7.91}    & 89.58 \text{\tiny ± 9.55}    \\
 &   & F + $\hat{\text{C}}$ & 37.50 \text{\tiny ± 7.91}    & 100.00 \text{\tiny ± 0.00}   & 56.25 \text{\tiny ± 11.86}   & 64.58 \text{\tiny ± 32.07}   \\ \hline
\multirow{12}{*}{\textbf{G-mean}} 
 &  & F           & 50.00 \text{\tiny ± 25.82}   & 25.00 \text{\tiny ± 22.36}   & 25.00 \text{\tiny ± 22.36}   & 33.33 \text{\tiny ± 14.43}   \\
 & a  & F + C       & 75.00 \text{\tiny ± 22.36}   & 75.00 \text{\tiny ± 22.36}   & 50.00 \text{\tiny ± 25.82}   & 66.67 \text{\tiny ± 14.43}   \\
 &   & F + $\hat{\text{C}}$ & 75.00 \text{\tiny ± 22.36}   & 50.00 \text{\tiny ± 25.82}   & 25.00 \text{\tiny ± 22.36}   & 50.00 \text{\tiny ± 25.00}   \\ \cline{2-7}
 &  & F           & 49.19 \text{\tiny ± 20.29}   & 57.12 \text{\tiny ± 16.36}   & 62.83 \text{\tiny ± 16.47}   & 56.38 \text{\tiny ± 6.85}    \\
 &  b & F + C       & 81.41 \text{\tiny ± 5.15}    & 65.08 \text{\tiny ± 17.07}   & 67.47 \text{\tiny ± 17.66}   & 71.32 \text{\tiny ± 8.82}    \\
 &   & F + $\hat{\text{C}}$ & 60.94 \text{\tiny ± 15.43}   & 31.02 \text{\tiny ± 13.06}   & 64.79 \text{\tiny ± 16.60}   & 52.25 \text{\tiny ± 18.49}   \\ \cline{2-7}
 &  & F           & 26.24 \text{\tiny ± 16.09}   & 36.13 \text{\tiny ± 15.83}   & 21.99 \text{\tiny ± 14.36}   & 28.12 \text{\tiny ± 7.26}    \\
 & c  & F + C       & 72.94 \text{\tiny ± 8.62}    & 76.46 \text{\tiny ± 9.20}    & 28.61 \text{\tiny ± 17.58}   & 59.34 \text{\tiny ± 26.67}   \\
 &   & F + $\hat{\text{C}}$ & 46.48 \text{\tiny ± 14.17}   & 47.91 \text{\tiny ± 13.87}   & 24.52 \text{\tiny ± 15.31}   & 39.64 \text{\tiny ± 13.11}   \\ \cline{2-7}
 &  & F           & 37.50 \text{\tiny ± 21.41}   & 55.18 \text{\tiny ± 18.84}   & 55.18 \text{\tiny ± 18.84}   & 49.29 \text{\tiny ± 10.21}   \\
 &  d & F + C       & 85.35 \text{\tiny ± 7.56}    & 75.00 \text{\tiny ± 22.36}   & 55.18 \text{\tiny ± 18.84}   & 71.84 \text{\tiny ± 15.33}   \\
 &   & F + $\hat{\text{C}}$ & 55.18 \text{\tiny ± 18.84}   & 92.68 \text{\tiny ± 6.55}    & 55.18 \text{\tiny ± 18.84}   & 67.68 \text{\tiny ± 21.65}   \\ \hline
\multirow{12}{*}{\textbf{MCC}} 
 &  & F           & 31.89 \text{\tiny ± 22.99}  & 38.15 \text{\tiny ± 17.48}  & 42.56 \text{\tiny ± 13.77}  & 37.53 \text{\tiny ± 5.36}   \\
 & b  & F + C       & 45.06 \text{\tiny ± 12.75}  & 52.62 \text{\tiny ± 16.90}  & 54.48 \text{\tiny ± 16.98}  & 50.72 \text{\tiny ± 4.99}   \\
 &   & F + $\hat{\text{C}}$ & 47.98 \text{\tiny ± 13.72}  & 11.02 \text{\tiny ± 6.77}   & 46.73 \text{\tiny ± 13.11}  & 35.24 \text{\tiny ± 20.99}  \\ \cline{2-7}
 &  & F           & 8.26 \text{\tiny ± 13.31}   & 17.49 \text{\tiny ± 19.71}  & 3.53 \text{\tiny ± 15.54}   & 9.76 \text{\tiny ± 7.10}    \\
 & c  & F + C       & 30.21 \text{\tiny ± 14.17}  & 34.96 \text{\tiny ± 15.84}  & 11.36 \text{\tiny ± 14.03}  & 25.51 \text{\tiny ± 12.48}  \\
 &   & F + $\hat{\text{C}}$ & 18.88 \text{\tiny ± 16.45}  & 22.53 \text{\tiny ± 14.87}  & 6.58 \text{\tiny ± 15.08}   & 16.00 \text{\tiny ± 8.36}   \\ \cline{2-7}
 &  & F            & -17.68 \text{\tiny ± 15.81} & 22.34 \text{\tiny ± 12.48}  & 14.44 \text{\tiny ± 12.91}  & 6.37 \text{\tiny ± 21.20}   \\
 & d  & F + C        & 30.25 \text{\tiny ± 15.65}  & 25.00 \text{\tiny ± 22.36}  & 22.34 \text{\tiny ± 12.48}  & 25.86 \text{\tiny ± 4.03}   \\
 &   & F + $\hat{\text{C}}$  & 14.44 \text{\tiny ± 12.91}  & 40.81 \text{\tiny ± 22.12}  & 14.44 \text{\tiny ± 12.91}  & 23.23 \text{\tiny ± 15.22}  \\ \hline 
\end{tabular}
}
\caption{Mean performance metric for different experimental settings varying the ACR breast density category. Each tabular shows the mean value together with the standard error calculated across the 5 cross-validation folds.}
\label{tab:acr}
\end{table}

\section{Conclusion}
We investigated the application of generative AI and multimodal deep learning to enable breast cancer virtual biopsy, addressing the challenge of missing CESM data in diagnostic workflows.
Our proposed approach integrates FFDM and CESM modalities in both CC and MLO views, utilizing CycleGAN to synthesize CESM images when real CESM data are unavailable.
Rather than centering on identifying the top-performing classifier, our study focused on demonstrating the efficacy of a multimodal framework capable of incorporating synthetic CESM images when required.
Our findings reveal that real CESM images significantly improve virtual biopsy performance compared to relying on FFDM alone.
Furthermore, when real CESM data is missing, synthetic CESM images generated by CycleGAN proved effective, particularly in multimodal configurations that combine FFDM and CESM modalities.
This ensures the continuity in the virtual biopsy process, even in the absence of real CESM data.
This study demonstrates the potential of generative AI to overcome limitations associated with missing imaging modalities, highlighting the value of integrating synthetic CESM images into multimodal virtual biopsy frameworks.
By addressing this challenge, the proposed approach has the potential to improve diagnostic workflows, providing clinicians with augmented intelligence tools that can help in diagnostic decision-making.
Additionally, we publicly release the dataset used in our experiments, which includes anonymized DICOM-format images along with clinical information derived from medical reports and biopsy results, aiming to support further research and innovation in this field.

Future research will explore the application of this methodology to larger and more diverse datasets, with a focus on developing the optimal neural network architecture for the virtual biopsy task.
Additionally, advancements in generative models and multimodal learning strategies will be pursued to further enhance the robustness and generalizability of the proposed framework.
Efforts will focus on improving the generation of CESM images in the MLO view through tailored preprocessing techniques or architectural enhancements to address view-specific complexities.
Moreover, we will investigate different multimodal fusion techniques, such as joint and early fusion, to optimize the integration of information from multiple modalities and further improve classification performance.
\label{sec:conclusion}

\section*{Author Contributions}
\textbf{Aurora Rofena}: Conceptualization, Data curation, Formal analysis, Investigation, Methodology, Software, Validation, Visualization, Writing – original draft, Writing – review and editing. \textbf{Claudia Lucia Piccolo}: Resources. \textbf{Bruno Beomonte Zobel}: Resources. \textbf{Paolo Soda}: Conceptualization, Funding acquisition, Methodology, Project administration, Supervision, Writing – review and editing. \textbf{Valerio Guarrasi}: Conceptualization, Data curation, Formal analysis, Investigation, Funding acquisition, Methodology, Project administration, Software, Supervision, Validation, Visualization, Writing – original draft, Writing – review and editing.



\section*{Acknowledgements}

Aurora Rofena is a Ph.D. student enrolled in the National Ph.D. in Artificial Intelligence, XXXVIII cycle, course on Health and life sciences, organized by Università Campus Bio-Medico di Roma.

Resources are provided by the National Academic Infrastructure for Supercomputing in Sweden (NAISS) and the Swedish National Infrastructure for Computing (SNIC) at Alvis @ C3SE, partially funded by the Swedish Research Council through grant agreements no. 2022-06725 and no. 2018-05973.
This work was partially founded by: i) PNRR MUR project PE0000013-FAIR, ii)  Cancerforskningsfonden Norrland project MP23-1122, iii) Kempe Foundation project JCSMK24-0094, iv) PNRR M6/C2 project PNRR-MCNT2-2023-12377755, v) Università Campus Bio-Medico di Roma under the program ``University Strategic Projects'' within the project ``AI-powered Digital Twin for next-generation lung cancEr cAre (IDEA)''.
\bb

%
%
\bibliographystyle{splncs04}
\bibliography{references.bib}

\clearpage  
\renewcommand{\thesection}{A\arabic{section}}
\renewcommand{\thetable}{A\arabic{table}}
\renewcommand{\thefigure}{A\arabic{figure}}

\section*{\centering Appendix}

\setcounter{section}{0}
\setcounter{table}{0}
\setcounter{figure}{0}

\begin{table}[h]
\centering
\setlength{\tabcolsep}{10pt} 
\begin{tabular}{lllll}
\toprule
\textbf{Classifier} & \textbf{Modalities} & \textbf{AUC} & \textbf{G-mean} & \textbf{MCC} \\ 
\midrule
\multirow{5}{*}{ResNet18} & F           & 61.52 \text{\tiny ± 10.67} & 51.29 \text{\tiny ± 13.04} & 16.19 \text{\tiny ± 13.43} \\
                          & C           & 88.57 \text{\tiny ± 5.00}  & 78.96 \text{\tiny ± 3.98}  & 61.76 \text{\tiny ± 6.48}  \\
                          & $\hat{\text{C}}$      & 72.72 \text{\tiny ± 4.75}  & 67.64 \text{\tiny ± 3.76}  & 34.05 \text{\tiny ± 6.33}  \\
                          & F + C       & 87.50 \text{\tiny ± 5.30}  & 76.55 \text{\tiny ± 3.91}  & 54.06 \text{\tiny ± 9.49}  \\
                          & F + $\hat{\text{C}}$  & 71.21 \text{\tiny ± 7.08}  & 68.18 \text{\tiny ± 5.17}  & 35.43 \text{\tiny ± 8.40}  \\ 
\midrule
\multirow{5}{*}{ResNet50} & F           & 60.84 \text{\tiny ± 4.34}  & 57.86 \text{\tiny ± 5.11}  & 25.31 \text{\tiny ± 12.16} \\
                          & C           & 87.47 \text{\tiny ± 5.37}  & 74.51 \text{\tiny ± 8.30}  & 51.95 \text{\tiny ± 12.32} \\
                          & $\hat{\text{C}}$      & 70.38 \text{\tiny ± 4.49}  & 59.32 \text{\tiny ± 7.20}  & 26.37 \text{\tiny ± 10.65} \\
                          & F + C       & 88.42 \text{\tiny ± 6.33}  & 76.95 \text{\tiny ± 7.17}  & 59.16 \text{\tiny ± 12.11} \\
                          & F + $\hat{\text{C}}$  & 73.66 \text{\tiny ± 4.58}  & 61.68 \text{\tiny ± 5.06}  & 28.58 \text{\tiny ± 7.56}  \\ 
\midrule
\multirow{5}{*}{VGG16}  & F           & 58.69 \text{\tiny ± 6.02}  & 54.43 \text{\tiny ± 5.99}  & 13.52 \text{\tiny ± 9.90}  \\
                          & C           & 74.45 \text{\tiny ± 5.19}  & 66.07 \text{\tiny ± 4.18}  & 37.68 \text{\tiny ± 9.77}  \\
                          & $\hat{\text{C}}$      & 49.98 \text{\tiny ± 5.62}  & 47.43 \text{\tiny ± 4.20}  & 5.62 \text{\tiny ± 10.90}  \\
                          & F + C       & 71.24 \text{\tiny ± 6.09}  & 64.83 \text{\tiny ± 4.99}  & 32.80 \text{\tiny ± 8.28}  \\
                          & F + $\hat{\text{C}}$  & 59.42 \text{\tiny ± 5.89}  & 59.20 \text{\tiny ± 4.77}  & 21.06 \text{\tiny ± 8.53}  \\ 
\bottomrule
\end{tabular}
\caption{Performance metrics for different classifiers and image combinations. Each tabular shows the mean value together with the standard error calculated across the 5 cross-validation folds.}
\label{tab:results}
\end{table}

\end{document}